\documentclass[aps,amsmath,amssymb,prl,twocolumn,showpacs]{revtex4}
\usepackage{epsfig}
\usepackage{graphicx}

\newcommand{\placefigure}[4]
{
 \begin{figure}[t]
 \includegraphics[width=#2]{#1}
 \vspace*{-.2cm}
 \caption{#3}
 \label{#4}
 \end{figure}
}

\begin{document}

\title{String Picture of a Frustrated Quantum Magnet and Dimer Model}
\author{Ying Jiang} 
\affiliation{Institut f\"ur Theoretische Physik, Universit\"at zu
  K\"oln, Z\"ulpicher Stra{\ss}e 77, 50937 K\"oln, Germany} 
\author{Thorsten Emig} 
\affiliation{Institut f\"ur
  Theoretische Physik, Universit\"at zu K\"oln, Z\"ulpicher Stra{\ss}e 77,
  50937 K\"oln, Germany}

\date{\today}

\begin{abstract}
  We map a geometrically frustrated Ising system with transversal
  field generated quantum dynamics to a strongly anisotropic lattice
  of non-crossing elastic strings. The combined effect of frustration,
  quantum and thermal spin fluctuations is explained in terms of a
  competition between intrinsic lattice pinning of strings and
  topological defects in the lattice. From this picture we obtain
  analytic results for correlations and the phase diagram which agree
  nicely with recent simulations.
\end{abstract}
\pacs{75.10.-b, 05.50.+q, 75.10.Nr}

\maketitle

The interplay of quantum and thermal fluctuations in geometrically
frustrated magnets raise new and fascinating issues both from a
theoretical and experimental perspective
\cite{Sachdev-book,Diep-book}.  Frustration prohibits the formation of
a collinear N{\'e}el state, inducing a macroscopic degeneracy of the
classical ground-state. Even for magnets with a discrete Ising
symmetry, the complexity of the ground-state manifold may endow the
system with a continuous symmetry.  Such symmetry is of particular
importance to 2D quantum magnets at low temperatures since then the
Mermin-Wagner theorem applies, precluding an ordered phase
\cite{Mermin+66}.  Another possible but contrary scenario is
``order-from-disorder'' where zero-point fluctuations select a small
particularly susceptible class of the ground-state manifold and yield
an ordered symmetry-broken state \cite{moessner1}.  This poses the
question if competing fluctuations about the classical ground-states
can lead to new strongly correlated states and (quantum) phase
transitions of unexpected universality classes.

The possibly simplest realization of classical frustration is provided
by the antiferromagnetic Ising model on a triangular lattice (TIAF).
This model is disordered even at zero temperature with a finite
entropy density and algebraic decaying spin correlations
\cite{Wannier+Houtappel}.  Quantum dynamics arise if a magnetic field
is applied transverse to the spin coupling. For this transverse field
TIAF, and its companions on other 2D lattices, Moessner and Sondhi
have argued the existence of both ordered and spin liquid phases
\cite{moessner1,Isakov+03}.  Experimental realizations of these models
can be found either directly in magnets with strong anisotropy, e.g.,
in LiHoF$_4$ \cite{Aeppli+98}, or indirectly (via a related (2+1)D
classical model) in stacked triangular lattice antiferromagnets
\cite{Collins+97} with strong couplings along the stack as studied in
recent experiments on CsCoBr$_3$ \cite{Mao+02}. Transverse field Ising
models also describe the singlet sector below the spin gap of
frustrated antiferromagnetic quantum Heisenberg models
\cite{Nikolic+03} and phases of quantum dimer models (QDM) on the dual
lattice whose ordering depends on the lattice structure
\cite{Moessner+01a+b}.  
Interestingly, a disordered QDM appears to be
a promising candidate for quantum computing \cite{Ioffe+02}.

In this Letter we explain the effect of simple quantum dynamics on a
frustrated Ising model by mapping the transversal field TIAF to a
stack of 2D lattices of non-crossing elastic strings.  We show that
the strings are described by a frustrated 3D XY model with a 6-fold
clock term. Since our mapping yields the XY coupling constants, we can
compute explicitly the phase diagram at {\it arbitrary} transverse
field strength. The diagram is shown in Fig.~\ref{fig:phasediagram}.
Previous studies of the model started from a Landau-Ginzburg-Wilson
(LGW) theory \cite{Blankschtein+84}, which neglects frustration
\cite{Coppersmith85}, to describe the vicinity of the quantum critical
point at strong transverse fields \cite{moessner1}.  If the phase of
the complex LGW order parameter is identified with the string
displacement, the LGW and string actions agree except for frustration.
However, the string approach shows that the configuration space is
restricted since strings cannot cross each other which constrains
their displacement. Moreover, the LGW approach does not provide the
coupling constants, including the sign of the clock-term which is
important for the selection of the ordered state.  Very recent Monte
Carlo simulations support the LGW prediction for the phase diagram but
the computations were performed for the related (2+1)D classical model
whose equivalence to the quantum problem involves a singular scaling
limit which makes simulations difficult \cite{Isakov+03}.  Recently,
the weak field behavior has been studied in terms of a quantum kink
crystal \cite{Mostovoy+03}.
\placefigure{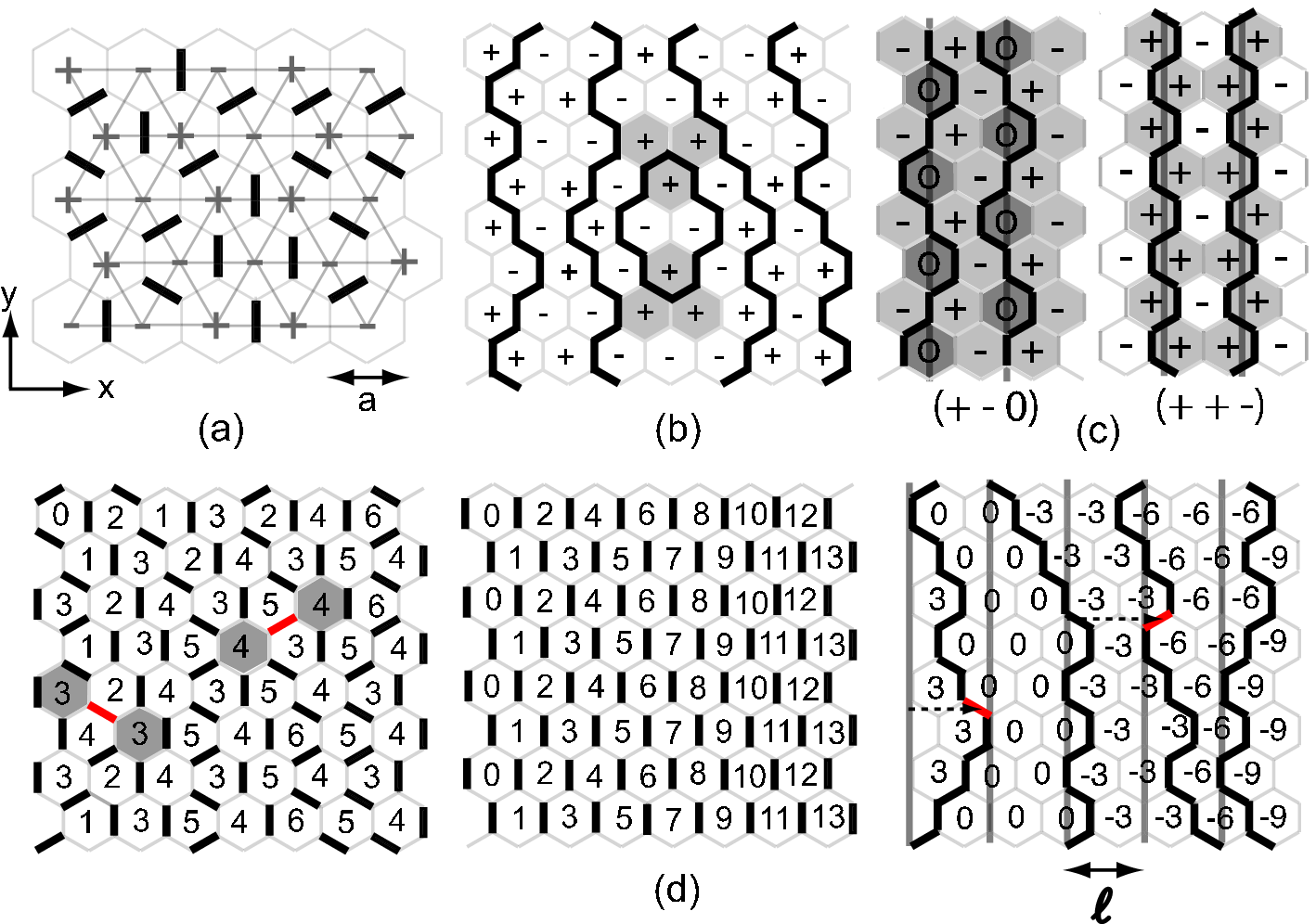}{0.95\linewidth}{(color online) (a) Relation
  between spins and dimers (b) Vortex--anti-vortex pair.  (c) The two
  flat states with flippable plaquettes (grey).  (d) Mapping of dimers
  to strings where the numbers denote the height profiles. The string
  displacement (arrows) is determined by the height of the two
  plaquettes (grey) which are joined by the displaced
  dimer.\vspace*{-.4cm}}{fig:spins+lines}

{\em Model for magnet, dimers, and strings.}---The transverse field TIAF
has the Hamiltonian
\begin{equation}
\label{eq:model}
H=J \sum_{\langle i,j \rangle} \,\sigma^z_i \sigma^z_j + 
\Gamma \sum_i \sigma^x_i \, ,
\end{equation}
with nearest neighbor coupling $J$, transverse field $\Gamma$, and
Pauli operators $\sigma^x$, $\sigma^z$.  First, we will consider the
classical ground states for $\Gamma=0$ which have one frustrated bond
per triangle.  A complete dimer covering of the dual hexagonal lattice
is obtained if a dimer is placed across each frustrated bond
\cite{Nienhuis+84}, see Fig.~\ref{fig:spins+lines}(a).  For each dimer
configuration a height profile $h$ can be defined on the triangular
lattice sites as follows.  Starting at an arbitrary site with some
integer number, one follows a triangle pointing down clockwise and
changes the height by +2 (-1) if a (no) dimer is crossed.  Repeating
this process for all down pointing triangles, one obtains a consistent
height on all sites, see Fig.~\ref{fig:spins+lines}(d). The string
representation follows from a given dimer state by the subtraction of
a fixed reference state with all vertical bonds occupied \cite{zeng1},
see Fig.~\ref{fig:spins+lines}(d). The strings fluctuate but remain
directed (along the reference direction) and non-crossing (due to the
frustration of the spin system).  They act as domain walls across
which the height (after subtraction of the reference height) changes
by $3$. A shift of $h$ by $3$ corresponds to a string translation by a
mean string separation $\ell=3a/2$ with $a$ the triangular lattice
constant.  Thus the string displacement is $u= \ell h/3 + u_0$ with
$h$ the coincident (original) height of two hexagonal plaquettes which
are joined by the displaced (non-vertical) string segment,
cf.~Fig.~\ref{fig:spins+lines}(d). $u_0$ is a global constant which
differentiates between two non-equivalent states of flat strings that
are not related by shifts by a lattice vector of the triangular
lattice, see Fig.~\ref{fig:spins+lines}(c).  For $u_0=n \, a/2$, $n
\in {\mathbb Z}$, the spins have orientations $(++-)$ on the three
sublattices and the non-vertical dimers sit on straight lines that are
locked {\it between} the triangular lattice sites. $u_0=a/4 + n \,
a/2$ selects orientations $(+-0)$ with the straight lines locked {\it
  on} the sites, i.e., one sublattice has alternating spins as
indicated in Fig.~\ref{fig:spins+lines}(c) by plaquettes with a zero.

By coarse-graining one obtains a continuous field $u({\bf r})$ which
allows to write the effective {\it free} energy of long-wavelength
fluctuations of the string lattice in the form of a continuum elastic
energy,
\begin{equation}
\label{eq:F-elastic}
{\cal F}_{\rm el}=\int d^2{\bf r} \left\{\frac {c_{11}}2(\partial_x u)^2 +
\frac{c_{44}}2(\partial_y u)^2 + V_L(u)\right\}
\end{equation}
with compression ($c_{11}$) and tilt ($c_{44}$) modulus which are both
of entropic origin.  $V_L(u)$ is a periodic potential which reflects
the discreteness of the lattice.  Since equivalent flat states are
related by shifts of all straight lines by $a/2$ and $u=a h/2+u_0$,
the locking potential must favor integer $h$,
\begin{equation}
\label{eq:lock-in-potential}
V_L = -v \cos(2\pi h) = -v \cos\left(4\pi (u-u_0)/a\right)
\end{equation}
with $v>0$ \cite{footnote1}.  For $\sqrt{c_{11}c_{44}}>2\pi/a^2$ the
potential $V_L$ is relevant (under renormalization) and the strings
lock into that state which maximizes the entropy by allowing for local
fluctuations that do not push the strings out of a tube of width
$\ell$ around their straight reference position, i.e., $|u| \le
\ell/2$. This condition restricts spin flips to certain plaquettes
[grey in Fig.~\ref{fig:spins+lines}(c)].  Taking into account that
strings are directed and non-intersecting, we find for a lattice of
$N$ sites that the states of orientation $(+-0)$ and $(++-)$ allow for
$2^{N/3+N/12}=\exp(0.2888 \, N)$ and $18^{N/12} \times 2^{(13/18)^3
  N/6} = \exp(0.2844\, N)$ localized configurations, respectively.
Thus $(+-0)$ is selected and $u_0=-a/4$ in
Eq.(\ref{eq:lock-in-potential}).

The classical spin correlations can be easily obtained within the
string picture. They follow from the local string density $\rho({\bf
  r})=\rho\,(1-\partial_x u(\bf r))$ with $\rho=1/\ell$ as
$\langle\sigma_i\sigma_j\rangle =\langle e^{i\pi\int_{x_j} ^{x_i}
  \rho({\bf r}) dx}\rangle$.  The compression modulus in
Eq.~(\ref{eq:F-elastic}) can be obtained from the equivalence of
non-crossing strings to 1D free Fermions \cite{Pokrovsky+79} which is
based on the Pauli principle and yields $c_{11}=\pi^2\rho^3/g$ with
$g=\rho c_{44}$ the string tension.  Then the periodic potential $V_L$
is irrelevant since $\sqrt{c_{11}c_{44}} = \pi \rho^2 < 2\pi/a^2$.
$g$ can be obtained from a random walk argument, however, its actual
value is unimportant since the mean squared displacement $\langle
[u({\bf r})-u({\bf 0})]^2\rangle = 1/(\pi \sqrt{c_{11}c_{44}}) \ln r$
becomes independent of $g$. This yields the known exact result
\cite{stephenson1}
\begin{equation}
\label{eq:spin-corr}
\langle\sigma_i\sigma_j\rangle=
\left(\frac{r_{ij}}{a}\right)^{-\eta}
\cos \frac{2\pi r_{ij}}{3a}, \quad \eta=\frac{1}{2} \, .
\end{equation}

{\em String model.}--- Now we consider a finite transversal
field. First, we use the Suzuki--Trotter theorem to express the
partition function of Eq.~(\ref{eq:model}) in terms of a $(2+1)$D {\it
  classical} Ising model with reduced Hamiltonian \cite{suzuki1}
\begin{equation}
\label{eq:3D-Ising}
H_{3D}=\sum_{<ij>,k=1}^n \tilde K_\| \sigma_{ik}\sigma_{jk} -
\sum_{i,k=1}^n \tilde K_\perp \sigma_{ik}\sigma_{ik+1}
\end{equation}
which consists of $n$ antiferromagnetic triangular lattices with
$\tilde K_\|=J/(nT)$ which are coupled ferromagnetically with $\tilde
K_\perp=\frac{1}{2}\ln(nT/\Gamma)$.  The correspondence becomes exact
for an infinite Trotter number $n \to \infty$. Next, we relate the
classical spins $\sigma_{ik}=\pm 1$ to the height profile $h_{ik}$
which is defined for each layer as above. The starting height is fixed
at one site in each layer by setting $h_{0k}=0$, ($3$) if
$\sigma_{0k}=+1$, ($-1$) along a column with layer index $k$.  Then
the height on all sites is fixed (modulo 6) for a given spin state
since parallel (anti-parallel) spins imply a change of $h$ by $+2$
($-1$) from site to site if the down pointing triangles are traversed
clockwise.  If one spin is flipped, $h$ changes on that site by $\pm
3$. One easily proofs that these rules are met by the relation
$\sigma_{ik} = \cos({\bf Q}{\bf R}_{ik}+\pi h_{ik}/3)=\pm 1$ with
${\bf Q}=(4\pi/(3a),0,0)$ and lattice sites ${\bf R}_{ik}$.  The
intra-layer coupling maps to
$-\cos[\frac{\pi}{3}(h_{ik}-h_{jk}+\eta_{ij})]$ with a shift
$\eta_{ij}=+1$, $-1$ for in-plane bond directions $(a,0)$,
$(a/2,\pm\sqrt{3}a/2)$, respectively.  For the inter-layer coupling
$\eta_{ij}=0$. In each layer the spin states map again to a string
lattice which, however, now contains vortex--anti-vortex pairs formed
by triangles with 3 frustrated bonds which span string loops, see
Fig.~\ref{fig:spins+lines}(b). Notice that the strings remain
non-crossing since a triangle can have either 1 or 3 frustrated bonds.
Including the lock-in potential of Eq.(\ref{eq:lock-in-potential})
with $u_0=-a/4$ and since $h=3(u-u_0)/\ell$, we obtain the reduced 3D
string Hamiltonian
\begin{eqnarray}
\label{eq:H-strings}
\!\!\!\!\!\!\!\!\!\!
H_S\!\!&=&\!\!-\tilde K_\|\!\!\sum_{<ij>,k}\!
\cos\left[\frac{\pi}{\ell}(u_{ik}-u_{jk}+\eta_{ij}a/2)\right]
\nonumber \\
&-&\!\!\tilde K_\perp\!\sum_{i,k}\!\cos\left[\frac{\pi}{\ell}
(u_{ik}\!-\!u_{ik+1})\right]
+v\!\sum_{i,k}\!\cos\left[\frac{6\pi}{\ell} u_{ik}\right]
\end{eqnarray}
with $v>0$ and the shift $\eta_{ij}$ reflecting frustration.  Since
$u_{ik}$ can vary over the bonds of a triangle only by $+a$ or $-a/2$,
the energy is minimized for a non-uniform change of $u_{ik}$ along the
triangles which is distinct from a continuous field whose oriented
uniform change defines a helicity, giving rise to a ${\mathbb Z}_2$
symmetry \cite{leedh1}. Thus the Ising model of Eq.(\ref{eq:3D-Ising})
maps to a stack of 2D lattices of {\it non-crossing} strings which is
described by a (2+1)D frustrated XY model with a 6-fold clock term.
This resembles the GLW theory \cite{Blankschtein+84} if the string
displacement is identified with the phase $\phi$ of the order
parameter via $\phi=\pi u/\ell$.  However, there are two important
differences. (i) the in-plane XY coupling is frustrated and (ii) there
is a topological constraint on $\phi$ since $u$ is restricted by the
non-crossing condition which increases the phase stiffness on large
length scales. The XY coupling allows for vortex loops which are in
general superpositions of two types. Vortex-anti-vortex pairs
[cf.~Fig.~\ref{fig:spins+lines}(b)] occurring in many layers form
loops oriented perpendicular to the planes, while the boundaries of 2D
regions along which strings in adjacent planes are shifted by $2\ell$
form parallel loops.  If the loop size is bound, the XY couplings of
Eq.~(\ref{eq:H-strings}) can be expanded in $u_{ik}$, and each layer
is described by Eq.~(\ref{eq:F-elastic}) with a harmonic interlayer
coupling which can render the lock-in potential relevant.

{\em Phase diagram.}--- First, we note that the layers cannot decouple
independently from the vortex unbinding transition in the layers since
parallel loops of diverging size can occur only at a critical value
for the effective $K_\|$ which is by factor of $1/8$ below that for
the in-plane dissociation of vortices \cite{Korshunov90}.  In
addition, a $p$-fold clock term is irrelevant under renormalization if
$p\gtrsim 3.4$ \cite{Aharony+86}, and thus we expect a quantum phase
transition in the 3D XY universality class.  To locate the transition,
we factorize the partition function of Eq.~(\ref{eq:H-strings}) into a
spin wave and a vortex part by the substitution $\tilde K
[\cos(\phi)-1] \to \sum_{m=-\infty}^\infty \exp[-K(\phi-2\pi m)^2/2]$,
denoted as Villain coupling \cite{jose1}. The couplings $K$ are known
in two limits, $K= \tilde K$ for $\tilde K\to \infty$ and $K =
1/(2\ln(2/\tilde K))$ for $\tilde K\to 0$. In fact, these limits are
realized for $\tilde K_\perp$ and $\tilde K_\|$ at large Trotter
numbers $n$, leading to
\begin{equation}
\label{eq:Villain-Ks}
K_\perp = \frac{1}{2} \ln(nT/\Gamma), \quad K_\|
= \frac{1}{2}\ln^{-1}(2nT/J).
\end{equation}
At $T=0$, one can set $T\sim J/n$ so that for $n \to \infty$ the
coupling $K_\perp$ remains finite, and the system shows 3D behavior.
For any finite $T$, however, $K_\perp$ must diverge with $n$ and the
system behaves effectively as 2D on large length scales.  Eliminating
$n$ from Eq.~(\ref{eq:Villain-Ks}), we obtain a relation between
$K_\|$ and $K_\perp$ which makes the parameter space 1D.  At large
$n$, we can expand this relation, giving
\begin{equation}
\label{eq:rel-between-Ks}
\sqrt{K_\| K_\perp} 
=\frac{1}{2} \left[1-\ln(2\Gamma/J)
\, K_\|\right] \, .
\end{equation}
For layered XY models, dimensional crossover scaling
\cite{Ambegaokar+80+Schneider} can be used to obtain a relation
between the 3D critical value $K_\infty^c$ for the in-plane coupling
and the corresponding $K_n^c$ for a system of $n$ layers,
\begin{equation}
\label{eq:dim-cross}
\frac{1}{n}\frac{K_\infty^{c}}{K_n^{c}}=\gamma
\left(\frac{K_\|}{K_{\perp}}\right)^{1/2}
\left(1-\frac{K_\infty^{c}}{K_n^{c}}\right)^\nu 
\end{equation}
with the 3D XY exponent $\nu\approx 2/3$ and a constant $\gamma$. This
yields for $n=1$ in the limit $K_\infty^c \ll K_1^c$ the following
relation between $K_\infty^c$ and $K_1^c$,
\begin{equation}
\label{eq:3d-k}
K_\infty^c = K_1^c \left( \nu + \gamma^{-1} \sqrt{K_\perp/K_\| }\right)^{-1} \, .
\end{equation}
At the quantum critical point the classical string system of
Eq.(\ref{eq:H-strings}) is 3D and thus at $K_\|=K_\infty^c$ the
relations of Eqs.~(\ref{eq:rel-between-Ks}) and (\ref{eq:3d-k}) must
be identical for consistency. This implies $\gamma=1/(2K_1^c)$ and
$\Gamma_c/J=\frac{1}{2} e^{\nu/K_1^c}$. The usual Kosterlitz-Thouless
(KT) argument yields for the vortex unbinding transition $K_1^c=2/\pi
\times 2/\sqrt{3}$ on the triangular lattice \cite{leedh1}, leading to
$\Gamma_c/J=1.24$.  However, this estimate neglects renormalization
effects due to the clock term, frustration and the non-crossing of
strings which should provide a net increase of $\Gamma_c$.
Interestingly, recent Monte Carlo studies suggest $\Gamma_c/J\approx
1.65$ \cite{Isakov+03}. For $\Gamma < \Gamma_c$ the clock term is
relevant, and the system is ordered.

\placefigure{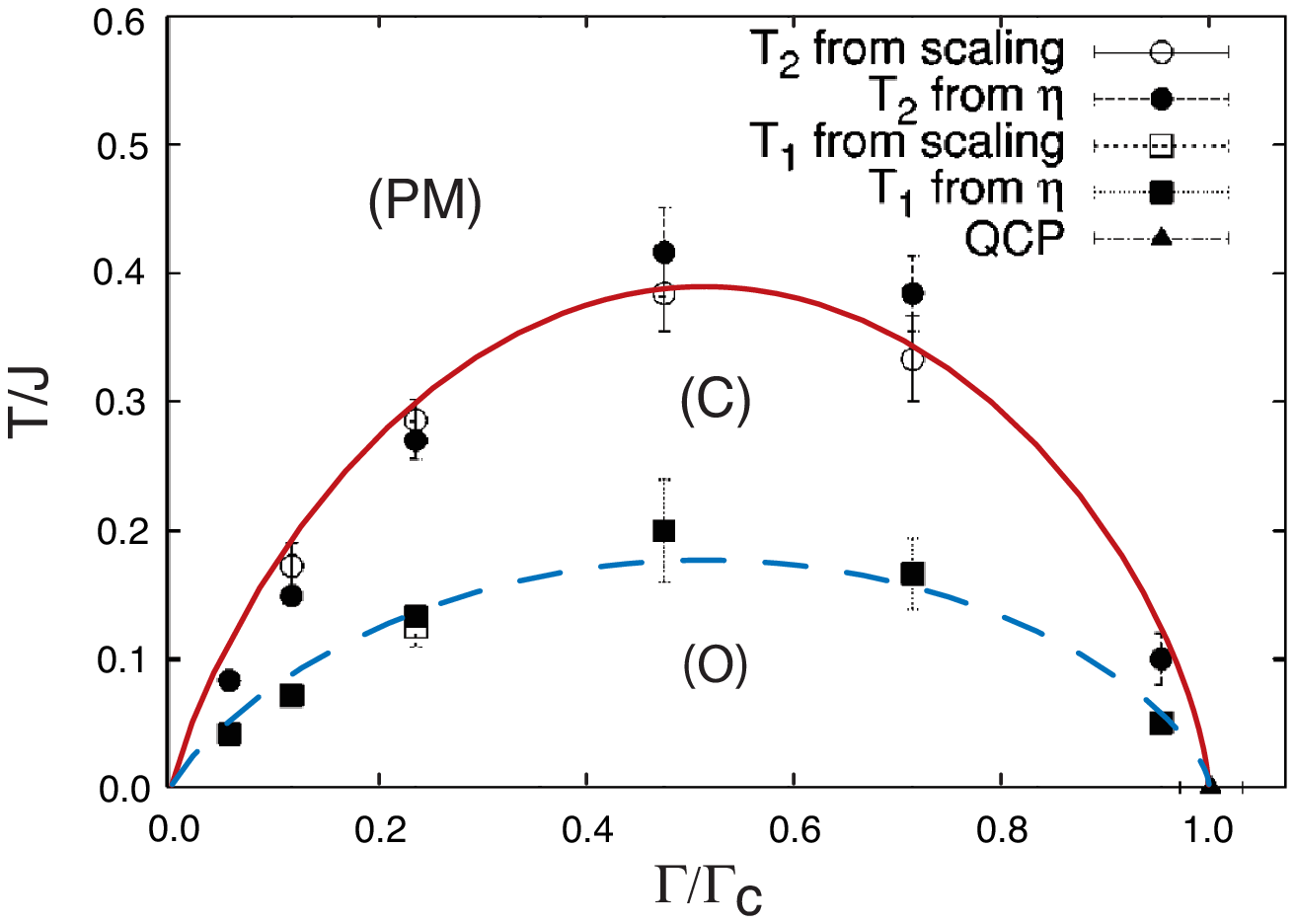}{0.95\linewidth}{(color online) Phase
  diagram predicted by Eq.~(\ref{eq:boundary}) and  Monte
  Carlo results of Fig.~1 in Ref. \cite{Isakov+03}. A critical phase
  (C) is separated by a boundary at $T_{c,1}$ from the paramagnetic
  phase (PM) and at $T_{c,2}=4/9\, T_{c,1}$ from an ordered phase (O).
  \vspace*{-.4cm}}{fig:phasediagram}

At finite temperature, 2D XY physics dominate at large scales, and KT
singularities are expected \cite{jose1}.  Using $K_\infty^c$ from
Eq.~(\ref{eq:3d-k}) in Eq.(\ref{eq:dim-cross}) with $K_\|=K_n^c$ we
obtain the critical coupling $K_n^c(K_\perp)$ as function of $K_\perp$
and large but finite $n$.  This result can be used to determine the
vortex unbinding transition at finite $T$.  We set $\zeta_0\equiv
e^{2K_\perp}$ so that $n=\zeta_0 \Gamma/T$ in $K_n^c(K_\perp)$.  The
renormalization of the coupling constants $K_\perp$, $K_\|$ must be
dependent, and from Eq.~(\ref{eq:Villain-Ks}) the effective $K_\|^{\rm
  eff}=1/(2\ln(2\zeta\Gamma/J))$ where $\zeta\equiv e^{2K_\perp^{\rm
    eff}}$ is the renormalized $\zeta_0$. Now we can use
$K_n^c=K_\|^{\rm eff}$ and $K_\perp=K_\perp^{\rm eff}$ and expand
$K_n^c(K_\perp)$ for large $\zeta$ which yields in the limit $n \to
\infty$ the phase boundary of a critical phase (C) with bound defects,
\begin{equation}
\label{eq:boundary}
\frac{T_{c,1}}{J} = b \, \frac{\Gamma}{\Gamma_c} \, \ln^\nu
\left( \frac{\Gamma_c}{\Gamma} \right)  \, ,
\end{equation}
where $b=\gamma e^{\nu/K_1^c}\zeta_0(2\ln\zeta)^{-5/3}$ is a numerical
constant which is fixed by the (unknown) renormalization of $K_\perp$
and remains finite for $n \to \infty$. However, for consistency, since
$\zeta_0=nT/\Gamma$ the effective coupling must behave at large $n$
like $K_\perp^{\rm eff} \sim (n/n_c)^{3/5}$ with a characteristic
$n_c=bK_1^c (J/T) (\Gamma/\Gamma_c)$ which is a measure for the
strength of quantum fluctuations and characterizes the effective
system size along the Trotter (``imaginary time'') axis.  At a lower
$T_{c,2}$ there is a second transition to an ordered state where the
clock-term in Eq.~(\ref{eq:H-strings}) becomes relevant and locks the
strings to the lattice. For a single layer, this transition is again
of KT type with critical $\hat K_1^c=9/(2\pi) \times 2/\sqrt{3}$
\cite{jose1}, and the boundary of the ordered phase can be obtained
analogously to $T_{c,1}$ but with $\gamma$ replaced by $\gamma
K^c_1/\hat K^c_1$ in $b$ below Eq.~(\ref{eq:boundary}), yielding
$T_{c,2}=(4/9) T_{c,1}$. Close to the quantum critical point both
temperatures vanish $\sim (\Gamma_c - \Gamma)^\nu$ as expected from
scaling. Fig.~\ref{fig:phasediagram} compares the result of
Eq.~(\ref{eq:boundary}) to recent Monte Carlo data for the phase
boundaries, showing very good agreement across the entire range of
$\Gamma$ for $b=0.98$.  The spin correlations in phase C decay
according to Eq.~(\ref{eq:spin-corr}) with $\eta$ varying continuously
between $\eta=1/4$ at $T_{c,1}$ and $\eta=1/9$ at $T_{c,2}$
\cite{jose1}. At the quantum critical point one has the 3D XY result
$\eta\approx 0.040$ \cite{Zinn-Justin}. The ordered phase (O) is
characterized by finite sublattice magnetizations
$(\sqrt{3}/2,-\sqrt{3}/2,0)$ which follows from $\sigma_{jk} =
\cos({\bf Q} {\bf R}_{jk} +2\pi(u_{jk}-u_0)/3) = \cos({\bf Q} {\bf
  R}_{jk} +\pi/6)$ for flat strings ($u_{jk}=0$) with $u_0=-a/4$ in
class $(+-0)$. This kind of order is consistent with simulations
\cite{Isakov+03}.

An exciting perspective is that the string analogy applies also to
frustrated Ising systems with general couplings and quenched disorder
examined in recent experiments \cite{Duijn+04}.  A glassy state of the
strings, similar to a vortex glass \cite{emig03+04}, would be an
interesting possibility.

We thank H.~Rieger and S.~Scheidl for helpful discussions. Support by
the Emmy Noether grant No.~EM70/2-3 and the SFB 608 from the DFG is
acknowledged.


\vspace*{-.25cm}
\begin{thebibliography}{99}
\vspace*{-.25cm}  

\bibitem{Sachdev-book} S. Sachdev, {\it Quantum Phase Transitions}
  (Cambridge University Press, Cambridge, 1999).
  
\bibitem{Diep-book} {\it Magnetic systems with competing interactions}, edited by
  H. T. Diep (World Scientific, Singapore, 1994)

\bibitem{Mermin+66} N. D. Mermin and H. Wagner, Phys. Rev. Lett. {\bf
    17}, 1133 (1966). 

\bibitem{moessner1} R. Moessner, S. L. Sondi and P. Chandra, Phys.
  Rev. Lett {\bf 84}, 4457 (2000); R. Moessner and S.L. Sondhi, Phys. Rev. B {\bf
    63}, 224401 (2001).

\bibitem{Wannier+Houtappel} G. H. Wannier, Phys. Rev. {\bf 79}, 357
  (1950); R. M. F. Houtappel, Physica (Amsterdam) {\bf 16}, 425
  (1950).

\bibitem{Isakov+03} S. V. Isakov and R. Moessner, Phys. Rev. B {\bf
    68}, 104409 (2003).
  
\bibitem{Aeppli+98} G. Aeppli and T. F. Rosenbaum, in {\it Dynamical
    Properties of Unconventional Magnetic Systems}, edited by A. R.
    Skjeltorp and D. Sherrington (Kluwer Academic, Amsterdam, 1998).

\bibitem{Collins+97} M. F. Collins and O. A. Petrenko, Can. J. Phys.
    {\bf 75}, 605 (1997).
    
\bibitem{Mao+02} M. Mao, B. D. Gaulin, R. B. Rogge, and Z. Tun,
    Phys. Rev. B {\bf 66}, 184432 (2002).

\bibitem{Nikolic+03} P. Nikolic and T. Senthil, Phys. Rev. B {\bf
      68}, 214415 (2003).

\bibitem{Moessner+01a+b} R. Moessner and S. L. Sondhi, Phys. Rev.
  Lett.  {\bf 86}, 1881 (2001); R. Moessner, S. L. Sondhi, and P.
  Chandra, Phys.  Rev. B {\bf 64}, 144416 (2001).
  
\bibitem{Ioffe+02} L. B. Ioffe {\it et al.}, Nature {\bf 415}, 503
    (2002).

\bibitem{Blankschtein+84} D. Blankschtein, M. Ma, A. N. Berker, G. S.
  Grest, and C. M. Soukoulis, Phys. Rev. B {\bf 29}, 5250 (1984).

\bibitem{Coppersmith85} S. N. Coppersmith, Phys. Rev. B {\bf 32}, 1584
  (1985).

\bibitem{Mostovoy+03} M.V. Mostovoy, D.I. Khomskii and J. Knoester, Phys. Rev.
Lett. {\bf 90}, 147203 (2003).

\bibitem{Nienhuis+84} B. Nienhuis, H. J. Hilhorst and H.  W. J.
  Bl\"ote, J. Phys. A: Math. Gen. {\bf 17}, 3559 (1984).

\bibitem{zeng1} C. Zeng, A. Middleton and Y. Shapir, Phys. Rev. Lett. {\bf 77},
3204 (1996).

\bibitem{footnote1} Since the height $h$ is defined modulo an
  arbitrary (non-integer) global constant $h_0$, the minima of
  $V_L(h)$ can be shifted by $h_0$.  However, $V_L(u)$ remains
  unchanged since the string positions are not effected (modulo $a/2$)
  by such a shift but then $u = \ell (h-h_0)/3 +u_0$.
 
\bibitem{Pokrovsky+79} V. L. Pokrovsky and A. L. Talapov, Phys. Rev.
  Lett. {\bf 42}, 65 (1979).

\bibitem{stephenson1} J. Stephenson, J. Math. Phys. {\bf 11}, 413 (1970).

\bibitem{suzuki1} M. Suzuki, Prog. Theor. Phys. {\bf 56}, 1454 (1976).

\bibitem{leedh1} D. H. Lee, J. D. Joannopoulos, J. W. Negele and
D. P. Landau, Phys. Rev. B {\bf 33}, 450 (1986).

\bibitem{Korshunov90} S. E. Korshunov, Europhys. Lett. {\bf 11}, 757
  (1990).

\bibitem{Aharony+86} A. Aharony {\it et al.}, Phys. Rev. Lett. {\bf
    57}, 1012 (1986).

\bibitem{jose1} J.V. Jos\'e, L.P. Kadanoff, S. Kirkpartrick and D.R. Nelson,
Phys. Rev. B {\bf 16}, 1217 (1977).

\bibitem{Ambegaokar+80+Schneider} V. Ambegaokar, B. I. Halperin, D. R.
  Nelson, and E. D. Siggia, Phys. Rev. B {\bf 21}, 1806 (1980); T.
  Schneider and A. Schmidt, Jour. Phys. Soc. Jap.  {\bf 61}, 2169
  (1992).

\bibitem{Zinn-Justin} J. Zinn-Justin, {\it Quantum field theory and
    Critical Phenomena} (Oxford University Press, New York, 1990).

\bibitem{Duijn+04} J. van Duijn {\it et al.}, Phys. Rev. Lett. {\bf
    92}, 077202 (2004).

\bibitem{emig03+04} T. Emig and S. Bogner, Phys. Rev. Lett. {\bf 90},
  185701 (2003); S. Bogner, T. Emig, A. Taha and C. Zeng, 
Phys. Rev. B {\bf 69}, 104420 (2004).

\end{thebibliography}
\end{document}